%% file: RTC-LCH_arxiv.tex
\documentclass[amsmath,amssymb,preprintnumbers,nofootinbib,prd,a4paper,twocolumn]{revtex4-1}
\pdfoutput=1
\usepackage{amsthm}
\usepackage{amsmath}
\usepackage{graphicx}
\usepackage{bbold}
\usepackage{color}
\usepackage{subfigure}
\usepackage{multirow}
\usepackage{graphicx,bm}
\usepackage{physics}

\usepackage{dcolumn}
\usepackage{bm}
\usepackage{slashed}
\usepackage{epsfig}
\usepackage[colorlinks=true, linkcolor=black, citecolor=black, urlcolor=black]{hyperref}
\usepackage{verbatim}
\usepackage[bordercolor=gray!20,backgroundcolor=blue!10,textsize=tiny]{todonotes}

\usepackage[utf8]{inputenc}
\usepackage[T1]{fontenc}
\usepackage[normalem]{ulem}

\newcommand{\beq}{\begin{eqnarray}}
\newcommand{\eeq}{\end{eqnarray}}

\newcommand{\bmp}{\noindent\begin{minipage}{16cm}}
\newcommand{\emp}{\end{minipage}\vskip 7mm} 

\newcommand{\be}{\begin{eqnarray}}
\newcommand{\ee}{\end{eqnarray}}

\newcommand{\SU}{\mbox{SU}}
\newcommand{\SO}{\mbox{SO}}
\newcommand{\SP}{\mbox{Sp}}
\newcommand{\UU}{\mbox{U}}

\begin{document}
\title{Techni-Composite Higgs models with (a)symmetric dark matter candidates}

\author{Giacomo {\sc Cacciapaglia}}
\email{g.cacciapaglia@ipnl.in2p3.fr}
\affiliation{Institut de Physique des 2 Infinis (IP2I),
	CNRS/IN2P3, UMR5822, 69622 Villeurbanne, France}
\affiliation{Universit\' e de Lyon, Universit\' e Claude Bernard Lyon 1, 69001 Lyon, France}

\author{Mads T. {\sc Frandsen}}
\email{frandsen@cp3.sdu.dk}
\affiliation{CP$^3$-Origins, University of Southern Denmark, Campusvej 55, DK-5230 Odense M, Denmark}
\author{Wei-Chih {\sc Huang}}
\email{huang@cp3.sdu.dk}
\affiliation{CP$^3$-Origins, University of Southern Denmark, Campusvej 55, DK-5230 Odense M, Denmark}
\author{Martin {\sc Rosenlyst}}
\email{rosenlyst@cp3.sdu.dk}
\affiliation{CP$^3$-Origins, University of Southern Denmark, Campusvej 55, DK-5230 Odense M, Denmark}

\author{Philip {\sc Sørensen}}
\email{philip.soerensen@desy.de}
\affiliation{II. Institute of Theoretical Physics, Universit\"{a}t  Hamburg, 22761 Hamburg, Germany}
\affiliation{Deutsches Elektronen-Synchrotron DESY, Notkestr. 85, 22607 Hamburg, Germany}

\preprint{DESY-21-196}


\begin{abstract}
	
We propose a novel class of composite models that feature both a technicolor and a composite Higgs vacuum limit, 
resulting in an asymmetric dark matter candidate. 
These Techni-Composite Higgs models are based on an extended left-right electroweak symmetry with a pseudo-Nambu Goldstone boson Higgs and stable dark matter candidates charged under a global $\mathrm{U}(1)_X$ symmetry, connected to the baryon asymmetry at high temperatures via the $\SU(2)_{\rm R}$ sphaleron.  
We consider, as explicit examples, four-dimensional gauge theories with fermions charged under a new confining gauge group  $G_{\rm HC} $.

\end{abstract}

\maketitle


The nature of Dark Matter (DM) and the origin of its relic density are arguably among the most important open questions in particle physics~\cite{Bertone:2018krk}. 
An answer could be found within existing solutions to other problems in nature:
for instance, we know that the bulk of the baryonic mass is due to strong dynamics while the baryonic relic density is due to a particle/anti-particle asymmetry. Furthermore, the baryonic and DM relic densities are of the same order. Within Technicolor (TC) models of the electroweak (EW) symmetry breaking~\cite{Weinberg:1975gm,Eichten:1979ah}, this observation has motivated DM candidates whose mass is due to new strong interactions and whose stability is due to $\UU(1)_X$ techni-baryon charge~\cite{Nussinov:1985xr,Barr:1990ca}. 
The baryon and DM relic densities can share a common asymmetric origin if this $\UU(1)_X$ global symmetry is anomalous under the EW sphalerons, paralleling the EW anomaly of the baryon number~\cite{Kuzmin:1985mm}. 
This realizes asymmetric dark matter (ADM)~\cite{Barr:1990ca} in TC models. In view of the progress at the LHC, the main drawback of TC models is that, in general, there is no simple parametric limit in which a Standard Model (SM)-like Higgs is recovered. 

One way out consists in engineering a vacuum misalignment of TC into a Composite Higgs (CH) model~\cite{Kaplan:1983fs,Dugan:1984hq,Contino:2003ve}, which is possible for a subset of TC theories~\cite{Cacciapaglia:2014uja,Cacciapaglia:2020kgq}. 
The composite Higgs state now arises as a pseudo-Nambu-Goldstone boson (pNGB) from the spontaneously broken (chiral) global symmetry. Hence, it can be parametrically close to the SM Higgs. 
However, in CH models the $\UU(1)_X$ techni-baryon or specie number is either broken or no longer anomalous under the EW sphalerons, thus the link between the baryon and DM asymmetric relic densities is severed (see~\cite{Frigerio:2012uc,Ma:2015gra,Ma:2017vzm,Ballesteros:2017xeg,Balkin:2017aep,Cai:2018tet,Balkin:2018tma,Cacciapaglia:2019ixa} for models with thermal composite DM). 

In this letter, we propose a new model building avenue where the CH and TC limits are simultaneously present, thus allowing for a successful description of the EW symmetry breaking and ADM. This requires:
\begin{itemize}
	\item[i)] a SM-like composite pNGB Higgs multiplet with custodial symmetry~\cite{Georgi:1984af}; 
	\item[ii)] a composite DM candidate, stable due to a $ \UU(1)_X $ symmetry of the strong interactions;
	\item[iii)] an EW anomaly of the $ \UU(1)_X $ symmetry allowing for shared asymmetry of baryons and DM;
	\item[iv)] a suppressed DM thermal relic density. 
\end{itemize}
The first ingredient i) is a key feature of CH models~\cite{Kaplan:1983fs}, holographic extra dimensions~\cite{Contino:2003ve,Hosotani:2005nz}, Little Higgs~\cite{ArkaniHamed:2001nc,ArkaniHamed:2002qx}, Twin Higgs~\cite{Chacko:2005pe} and elementary Goldstone Higgs models~\cite{Alanne:2014kea}. Extensions of the global symmetries can accommodate ii), 
however these model types typically do not satisfy ii) and iii) together.

To realize all four requirements of the Techni-Composite Higgs scenario, we are led to consider an extended Left-Right EW sector, with gauged $\SU(2)_{\rm L}\otimes \SU(2)_{\rm R}\otimes  \UU(1)_{\rm Y'}$ symmetry, dynamically broken by the strong dynamics via a left-right coset:  $G/H \supset G_{L}/H_{L} \otimes  G_{R}/H_{R}$. The left sub-coset $G_{L}/H_{L} $ is pinned in a CH direction by appropriate interactions, while the right sub-coset $G_{R}/H_{R}$ is in the TC vacuum with an unbroken global $\UU(1)_X$. This symmetry is anomalous under $\SU(2)_{\rm R}$, and the lightest composite state carrying $ X $-charge plays the role of ADM. Imposing the requirements i)-iv) also constrains non-trivially the form of the operators that generate the SM-fermion masses. Note that the interplay between TC and CH limits was used in Ref.~\cite{Cai:2019cow}, where the TC vacuum is only present at high temperatures. In our paradigm, the two limits coexist at all temperatures.

For concreteness, we consider four-dimensional gauge theories with a single strongly interacting gauge group $G_{\rm HC}$, with hyper-fermions that generate the global symmetry breaking via condensation. For models with a single fermionic representation,  the symmetry breaking patterns are known~\cite{Witten:1983tx,Kosower:1984aw}: Given $N$ Weyl spinors transforming as the HC $\mathcal{R}$ representation, the three possible classes of vacuum cosets are $ \SU(N)/\SO(N) $ for real $\mathcal{R}$, $ \SU(N)/\SP(N) $ for pseudo-real $\mathcal{R}$ and $ \SU(N)\otimes \SU(N)\otimes \UU(1)/ \SU(N)\otimes \UU(1)$ for complex $R$~\cite{Peskin:1980gc}.
The minimal CH cosets that fulfil requirement i), within these three classes, contain $N=5$ in the real case~\cite{Dugan:1984hq}, $N=4$ in both the pseudo-real~\cite{Galloway:2010bp} and the complex cases~\cite{Ma:2015gra}. In terms of pNGB spectrum, the pseudo-real case is the most minimal, with only 5 states.
Similarly, the minimal TC cosets that fulfil the requirements ii)-iv) contain $N=4$ in the real case~\cite{Sannino:2004qp,Dietrich:2006cm,Foadi:2007ue,Frandsen:2009mi}, $N=4$ in the pseudo-real~\cite{Ryttov:2008xe} and $N=2$ in the complex one~\cite{Kaplan:1983fs}. 
In the first two, the ADM candidates are pNGBs, while in the complex case it is a baryon~\cite{Kaplan:1983fs}.

To realize our scenario, we need to introduce two sets $\mathbb{S}_{L,R}$ of hyper-fermions, charged under $\SU(2)_{\rm L,R}$, respectively. The representations $\mathcal{R}_{L,R}$ may be different or identical, and they determine the resulting coset structure. We will assume that the pattern of the $ L $ and $ R $ cosets are the same as above even when the two representations are different, while for $\mathcal{R}_L \equiv \mathcal{R}_R$ the coset is enlarged.
The strong $G_{\rm HC}$ interactions produce condensates of the $\mathbb{S}_{L,R}$ fermions at scales $f_{L,R}$, which are of similar size. 
Hence, the breaking of the EW gauge symmetry occurs as follows:
\begin{equation}\begin{aligned}
&\SU(2)_{\rm L}\otimes \SU(2)_{\rm R}\otimes \UU(1)_{\rm Y'} \xrightarrow{f_R} \SU(2)_{\rm L}\otimes \UU(1)_{\rm Y}  
\\&\xrightarrow{f_L \sin\theta_L = v_{\rm EW}} \UU(1)_{\rm EM} \,,
\label{Eq:EWbreaking}
\end{aligned}
\end{equation}
where the hierarchy between the EW scale $v_{\rm EW}$ and the compositeness scale $f_L \sim f_R$ is generated in the $ \mathbb{S}_L $-sector and is parametrized by a (small) angle $\theta_L$.

The most minimal choice for the $\mathbb{S}_L$ sector consists in four Weyl spinors $Q_L$, arranged in one $\SU(2)_{\rm L}$ doublet $(U,D)$ and two singlets $\widetilde{U}$ and $ \widetilde{D}$, transforming as a pseudo-real representation $\mathcal{R}_L$ of $G_{\rm HC}$ and as a fundamental of a global $G_L = \SU(4)_L$. The minimal $ L $ coset will, therefore, contain the longitudinal components of the $W^\pm$ and $Z$ bosons, a Higgs candidate and a singlet $\eta$. We will focus on this scenario in the following, as shown in Table~\ref{table:fullmodel}.\footnote{The same gauge charge assignment can be used for a complex $\mathcal{R}_L$, at the price of including right-handed hyper-fermions with the same quantum numbers. For real $\mathcal{R}_L$ one would need two $\SU(2)_{\rm L}$ doublets with opposite $\UU(1)_{\rm Y'}$ charges and a neutral singlet, thus 5 Weyl spinors in total.}
The minimal $\mathbb{S}_R$ also contains four Weyl spinors $Q_R$, arranged in an $\SU(2)_{\rm R}$ doublet $(C,S)$ and two singlets $\widetilde{C}$ and $ \widetilde{S}$. The quantum numbers, together with the $\UU(1)_X$ charges, are shown in Table~\ref{table:fullmodel}, and are valid for all possible representations $\mathcal{R}_R$: in the complex case, however, the singlets $\widetilde{C}$ and $\widetilde{S}$ transform as the conjugate $\mathcal{R}_R^\ast$, while $\UU(1)_X$ is the techni-baryon number. 
In the pseudo-real case, the coset is $\SU(4)_{ R}/\SP(4)_{ R}$ and, besides the longitudinal modes of the $W_{\rm R}^\pm$ and $Z_{\rm R}$ bosons, the spectrum contains a complex neutral pNGB carrying $ X $-charge.  
In the real case, the coset $\SU(4)_{R}/\SO(4)_{ R}$ contains a complex neutral pNGB and two charged ones carrying $ X $-charge. 
In the complex case, the coset $\SU(2)_{ R1} \otimes \SU(2)_{ R2} \otimes \UU(1)_X/\SU(2)_{VR} \otimes \UU(1)_X$ does not contain pNGBs carrying $ X $-charge, so the DM candidate is played by a baryon-like state.
In all cases, the $\UU(1)_X$ symmetry has a gauge anomaly with respect to the $\SU(2)_{\rm R} \otimes \UU(1)_{\rm Y'}$ symmetry.
Note that, if $\mathcal{R}_L \neq \mathcal{R}_R$, there exists a global $\UU(1)_\Theta$ symmetry under which both sets of hyper-fermions are charged, which is spontaneously broken by the condensates and generates a light singlet pNGB, along the lines of Refs~\cite{Ferretti:2016upr,Belyaev:2016ftv}.
Finally, if $\mathcal{R}_L = \mathcal{R}_R \equiv\mathcal{R}$, the coset is enhanced: for pseudo-real $\mathcal{R}$, the coset is $\SU(8)/\SP(8)$; for real $\mathcal{R}$, the minimal case is $\SU(9)/\SO(9)$; for complex $\mathcal{R}$, we have $\SU(6)_1 \otimes \SU(6)_2 \otimes \UU(1)_{TB}/\SU(6)_{V} \otimes \UU(1)_{TB}$, where the $\UU(1)_{X}$ in Table~\ref{table:fullmodel} is a linear combination of $\UU(1)_{TB}$ and a $\UU(1)$ factor inside $\SU(6)_{V}$.

\begin{table}[t]
	\centering
	\begin{tabular}{ccccccc}
		\hline
		&$G_{\rm HC}$	&$\SU(3)_{\rm QCD}$	&$\SU(2)_{\rm L}$	&$\SU(2)_{\rm R}$	&$ \UU(1)_{\rm Y'} $	&$\UU(1)_X$\\
		\hline
		$ (U,D) $	&$\mathcal{R}_{L}$			&1			&$ \Box $		&1			&0				& 0\\
		$ \widetilde{U} $	&$\mathcal{R}_{L}$				&1			&1			&1			&$ -1/2	 $		& 0	\\
		$ \widetilde{D} $	&$\mathcal{R}_{L}$				&1		&1			&1			&$ +1/2 $			& 0	\\
		$ (C,S) $	&$\mathcal{R}_{R}$				&1				&1			&$ \Box $		&0				&$ +1$	\\
		$ \widetilde{C} $	&$\mathcal{R}_{R}$				&1			&1			&1			&$ -1/2 $			&$ -1$	\\
		$ \widetilde{S} $	&$\mathcal{R}_{R}$				&1			&1			&1			&$ +1/2 $			&$ -1$	\\ \hline
		$ q_{{\rm L},i} $			&1			&$ \Box $			&$ \Box $		&1			&$ +1/6 $		&0	\\
		$ q_{{\rm R},i} $			&1			&$ \Box $				&1			&$ \Box $		&$ +1/6 $		&0	\\
		$ l_{{\rm L},i} $			&1				&1				&$ \Box $		&1			&$ -1/2 $		&0	\\
		$ l_{{\rm R},i} $			&1				&1				&1			&$ \Box $		&$ -1/2 $		&0	\\
		\hline
	\end{tabular}
	\caption{Fermion field content and their charges of the Techni-Composite Higgs template models with the full left-right gauge symmetry. All groups are gauged except for $ \UU(1)_X $, which is a global symmetry in the $ R $ sector responsible for dark matter stability.}
	\label{table:fullmodel}
\end{table}

The remaining important ingredient for model building is the list of operators that generate the SM-fermion masses. The operators play an important role in determining the vacuum alignment in both $ L $ and $ R $ sectors, particularly via the top mass~\cite{Panico:2015jxa}.  To connect the $\UU(1)_X$ anomaly to the baryon number via $\SU(2)_{\rm R}$, we require that the right-handed SM fermions transform as doublets, as shown in the bottom rows of Table~\ref{table:fullmodel}.
The operators that generate the fermion masses, therefore, appear as 6-fermion operators with the generic structure $Q_L Q_L Q_R Q_R \psi_{\rm L} \psi_{\rm R}$. For instance, for the top quark
\begin{equation} \label{eq:topmass}
\frac{\xi_t}{\Lambda_t^5} (Q_L^T P_L Q_L)(Q_R^T P_R Q_R) q_{{\rm L},3} q_{{\rm R},3}  + {\rm h.c.}\,,
\end{equation}
where $P_L$ and $P_R$ are two-index matrices in the $G_L$ and $G_R$ space, respectively, selecting the appropriate combinations of the hyper-fermions that ensure gauge invariance and couple the top fields with the components that acquire a non-zero condensate. As such, $P_{L,R}$ transform as doublets of $\SU(2)_{\rm L,R}$, respectively. Similar operators can be added for all SM fermions. Note also that, for complex $\mathcal{R}_{L,R}$, it suffices to replace $Q_{L,R}^T$ by the conjugate hyper-fermions. We also remark that the operator in Eq.~\eqref{eq:topmass}, reminiscent of the mass terms in the original TC models~\cite{Eichten:1979ah}, can be generated via partial compositeness~\cite{Kaplan:1991dc} in models proposed in Refs~\cite{Barnard:2013zea,Ferretti:2013kya} via operators of the form:
\begin{equation} \begin{aligned}\label{eq: PC operator top yukawa}
\frac{y_L}{\Lambda_{t}}q_{{\rm L},3} (Q_L^T P_{L} Q_L \chi_t)+ \frac{y_R}{\Lambda_{t}}q_{{\rm R},3} (Q_R^T P_{R} Q_R \chi_t^\dagger)+{\rm h.c.}\,,
\end{aligned} \end{equation}
where $\chi_t$ is a new hyper-fermion, transforming in a suitable representation of $ G_{\rm HC}$, and carrying appropriate quantum numbers under the SM gauge symmetry. 
To illustrate the new scenario, in the following we will focus on a specific minimal model, and leave other examples to the Supplementary Material. \\

The minimal scenario we illustrate here is based on $G_{\rm HC} = \SP(6)_{\rm HC}$ with $\mathcal{R}_L = F$ (fundamental, pseudo-real) and $\mathcal{R}_R = A$ (antisymmetric, real). We will also include masses for the hyper-fermions in $\mathbb{S}_L$, because they help stabilize the CH vacuum~\cite{Cacciapaglia:2014uja,Dong:2020eqy}. The relevant physics can be described below the condensation scale in terms of an effective theory, following the CCWZ prescription~\cite{Marzocca:2012zn}, and based uniquely on the coset symmetry:
\begin{equation}
\frac{\SU(4)_{ L} \times \SU(4)_{ R} \times \UU(1)_\Theta}{\SP(4)_{ L} \times \SO(4)_{ R}}\,.
\end{equation}
At lowest order, the effective Lagrangian has the form
\begin{equation} \label{eq:Left}
\mathcal{L}_{\rm EFT} = \mathcal{L}_{\chi pt} - V_{\rm eff}\,,
\end{equation}
where the first term corresponds to the usual chiral perturbation theory for the pNGBs, and the second term contains the effective potential generated by loops of the SM fields, namely the EW gauge bosons and the fermions (top). The latter plays a crucial role in determining the vacuum alignment and the gauge symmetry breaking (see the Supplementary Material).

To investigate the asymmetric relic, we analyse the dynamics of the sphalerons associated with the left and right gauge symmetries.
The $\SU(2)_{\rm L} \times \SU(2)_{\rm R}$ sphaleron equations, when in equilibrium above $f_{L,R}$, yield chemical potential equations of the form
\begin{align}
\begin{aligned}
&\left( \mu_{u{\rm L},i} + 2\mu_{d{\rm L},i} \right) + \mu_{\nu {\rm L},i}+ \frac{d(\mathcal{R}_L)}{2}(\mu_{U}+\mu_{D}) = 0, \\
&\left( \mu_{u{\rm R},i} + 2\mu_{d{\rm R},i} \right) +  \mu_{\nu {\rm R},i}+ \frac{d(\mathcal{R}_R)}{2}\left(\mu_{C}+\mu_{S}\right) =0,\\
\end{aligned}\label{Sphaleron equilibirum}
\end{align}
where sums over the generations are left understood and $ d(\mathcal{R})$ is the dimension of representation $\mathcal{R}$. 
The labelling of the chemical potentials $ \mu $ follows Ref.~\cite{Ryttov:2008xe}. The alignment of the $\mathbb{S}_{L}$ sector vacuum also results in the separate equation $\mu_{U}+\mu_{D} = 0$. 
Together with equilibrium conditions from the 6-fermion operators in Eq.~\eqref{eq:topmass} and conditions on the relevant charges, these sphaleron processes yield a system that can be solved for the relic density of $ X $-charged states after condensation (see the Supplementary Material).

If all three families of SM fermions are gauged under $ \SU(2)_{\rm R}$ and all the operators, including both sphalerons, are in equilibrium, then the total $\UU(1)_X$ asymmetry is zero. However, due to the high scale of $f_{L,R}$ not all families are in equilibrium. A fermion $\psi$, receiving its mass $m_\psi$ from a Yukawa interaction generated by a 6-fermion operator such as Eq.~\eqref{eq:topmass}, is in equilibrium at a temperature $ T $ if~\cite{Davidson:2008bu, Deppisch:2017ecm}
\begin{gather}
2.3\times 10^4\times \sqrt{\frac{10\text{ TeV}}{f}}\frac{m_\psi}{v_{\rm EW}} \gtrsim \left(\frac{f}{T}\right)^{9/2}.\label{eq:eq-condition}
\end{gather}
The system will have non-trivial solutions if the 6-fermion operators are inefficient for at least one charged fermion but efficient for at least one other charged fermion.
By evaluating Eq.~\eqref{eq:eq-condition} at the condensation temperature $ T=f $, until which  expect the sphalerons to be active, we find this to be realised for 20 GeV $ \lesssim f \lesssim 3\times10^{12} $ GeV.

In this window, 6-fermion operators are inefficient for at least the electron but efficient for at least the top quark. 
Solving the set of equilibrium under these conditions we find, for both first- and second-order phase transitions, the following ratio:
\begin{align}
\begin{aligned}
\left\vert\frac{X}{B}\right\vert=&2\left(3+\frac{L}{B}\right),\label{eq:Upsilon}
\end{aligned}
\end{align} where $ X $, $ B $ and $ L $ are the $ X $-charge, lepton, baryon number densities respectively, and where we leave $ L $ and $ B $ as free parameters in this work. For second-order phase transitions, Eq.~(\ref{eq:Upsilon}) applies even when all 6-fermion operators are inefficient. For first-order phase transitions, the system is under-constrained in absence of efficient 6-fermion operators.
The ADM relic density can be expressed in terms of the charge densities as
\begin{gather} \label{eq: OmegaDM/OmegaB fundamental} 
\frac{\Omega_{DM}}{\Omega_B} = 
\left\vert\frac{X}{B}\right\vert \frac{m_{DM}}{2 m_p}\sigma\left(\frac{m_{DM}}{2 T_F}\right), 
\end{gather} 
where $T_F$ is the temperature of the phase transition
and $\sigma (x)$ is the Boltzmann suppression factor \cite{Harvey:1990qw,Ryttov:2008xe}. 

In absence of Boltzmann suppression ($m_{DM} \ll T_F$), the asymmetry sharing naturally fixes the DM number density to the order of the baryonic number density, so that $ \Omega_{DM}/\Omega_B  \sim \mathcal{O}(1)$ is found for $ m_{DM}\sim \mathcal{O}(m_p) $.
The correct relic density $\Omega_{DM}/\Omega_B = 5.36$ \cite{Planck:2018vyg}, therefore, requires either a small DM mass $m_{DM}\ll f_R$, a tuning in $\vert X/B\vert \ll 1$, or an exponential suppression in the Boltzmann factor for $T_F \ll m_{DM} $.  Another possibility is to allow the decay of the heavy $ X $-charged pNGB into another light stable state~\cite{Frandsen:2011kt}.

To assure condition iv), the DM pNGB needs to efficiently annihilate to suppress the thermal component of the relic density.
For heavy DM candidates ($ m_{DM} > m_{W_{\rm R}} = g_{\rm R} f_R/2 $), the dominant annihilation channel is in a pair of $ \SU(2)_{\rm R} $ gauge bosons, which has a cross section of the form:
\begin{equation} \begin{aligned}\label{Eq: sigma v DM DM to WR WR}
\langle  \sigma v\rangle_{W_{\rm R}} =&  \frac{g_{\rm R}^4 m_{DM}^2}{32\pi  m_{W_{\rm R}}^4}\sqrt{1-\frac{m_{W_{\rm R}}^2}{m_{DM}^2}}
\\ & \quad \quad\times\left(1-\frac{m_{W_{\rm R}}^2}{m_{DM}^2}+\frac{3}{4} \frac{m_{W_{\rm R}}^4}{m_{DM}^4}\right)\,.
\end{aligned}\end{equation} 
This channel is effective in wiping out the thermal relic if $ \langle  \sigma v\rangle_{W_{\rm R}} \gg 3\times 10^{-26}~\text{cm}^3\text{s}^{-1}$ \cite{Steigman:2012nb}, implying 
\begin{equation} \label{DMmass cond1}
m_{DM} \gg 0.073~\text{TeV} \times \left(\frac{f_R}{\text{TeV}}\right)^2\,,
\end{equation}
for $m_{DM} \gg m_{W_{\rm R}}$.
For lighter DM masses, $m_{DM} < m_{W_{\rm R}}$, the main annihilation mode involves a pair of pNGBs that do not carry $ X $-charges: by studying the potential, we found that the dominant channel involves the $\UU(1)_\Theta$ pNGB $\Theta$, which can be parametrically lighter than the other pNGBs~\cite{Belyaev:2016ftv}. 
The annihilation cross section reads
\begin{equation} \begin{aligned}\label{Eq: sigma v DM DM to eta eta}
&\langle  \sigma v\rangle_{\Theta} = \frac{\lambda_{XX\Theta\Theta}^2}{32\pi  m_{DM}^2}\sqrt{1-\frac{m_{\Theta}^2}{m_{DM}^2}}\,, 
\end{aligned}\end{equation}
where the quartic coupling is suppressed by the misalignment in the $ L $-coset, $\lambda_{XX\Theta\Theta} \sim \lambda_0 (v_{\rm SM}/f_L)^2$. This process can wipe out the thermal density for 
\begin{equation} \label{DMmass cond2}
m_{DM} \ll 0.21~\text{TeV} \times \left(\frac{\text{TeV}}{f_L}\right)^2 \lambda_0\,,
\end{equation}
for $m_\Theta \ll m_{DM}$.

\begin{figure}[t!]
	\centering
	\includegraphics[width=0.45\textwidth]{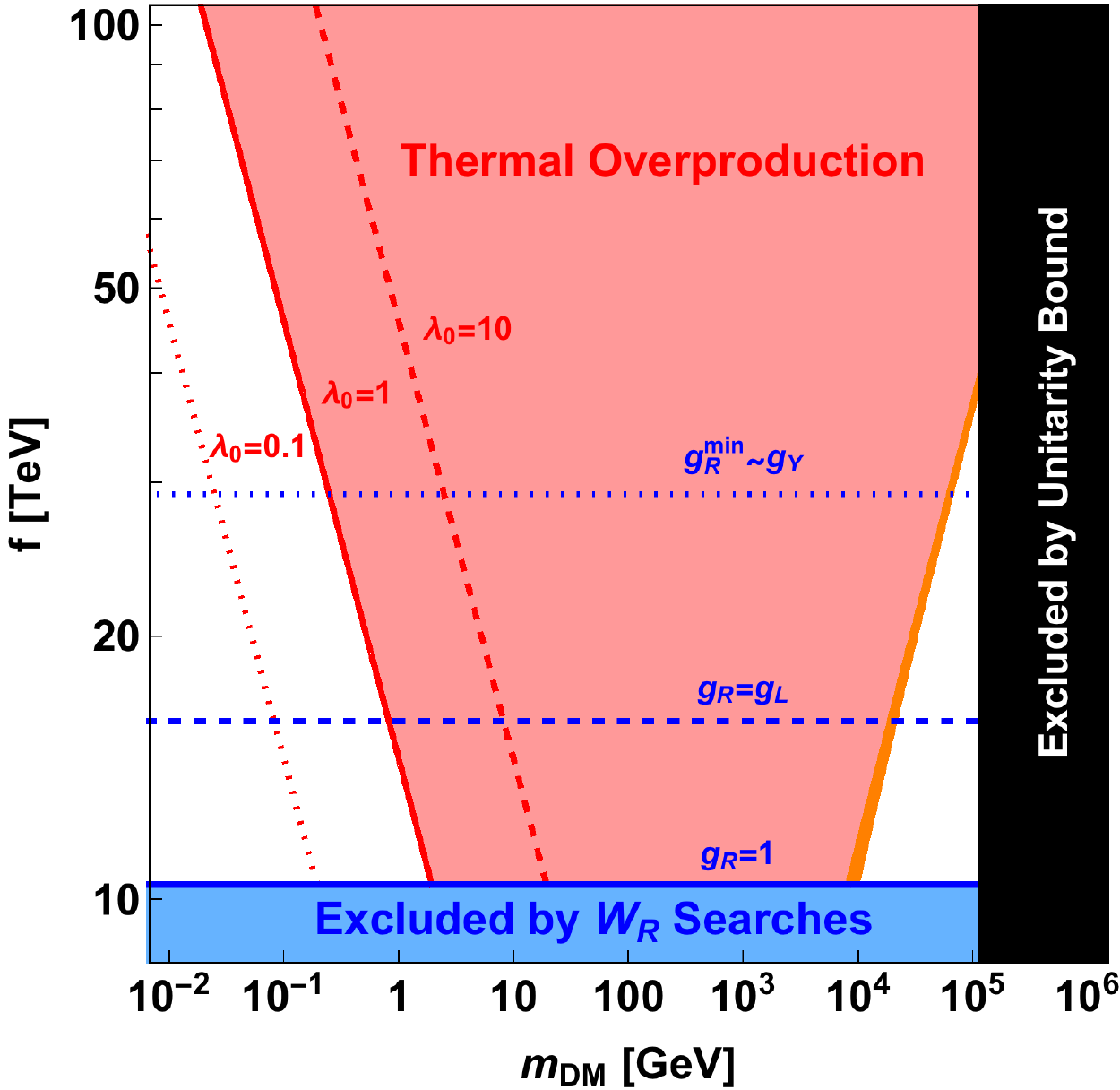} 
	\caption{Constraints on $ m_{DM} $ and the compositeness scale $ f\equiv f_L=f_R $. The red shaded region is excluded by thermal overproduction, while an upper limit on $m_{DM}\lesssim 110 $~TeV comes from unitarity bounds~\cite{Baldes:2017gzw}. The low mass limit of the excluded region depends on $\lambda_0$, and we show three sample values. A lower limit on $f$ comes from direct searches of $W_{\rm R}$ at colliders, which crucially depends on the value of $g_{\rm R}$ (we show three sample values: $ g_{\rm R}=1,\ g_{\rm L}, \ g_{\rm Y}$, with the last being the minimum allowed value). The high $m_{DM} $ edge of the red region has a mild dependence on $ g_{\rm R} $, illustrated by the thin orange band, moving to the right for larger $g_{\rm R}$.}
	\label{fig: Model Constraints}
\end{figure}

The strongest bound on the compositeness scale comes from direct searches at colliders for $W_{\rm R}$, as this state can be produced via Drell-Yann if it couples to the first generation. The most recent CMS bound from di-jet resonant searches~\cite{Sirunyan:2018mpc} reads  $ m_{W_{\rm R}}\gtrsim 5.2 $~TeV, which implies for $g_{\rm R} = g_{\rm L}$ a bound $f_R \gtrsim 16.4 $~TeV.
If $f_L = f_R$, this bound also implies a bound on the misalignment in the EW sector ($ L $-coset), which we can best express in terms of the fine tuning parameter $\xi = v_{\rm SM}^2/f_L^2$~\cite{Contino:2010rs}: the $W_{\rm R}$ searches imply a bound $\xi \lesssim 2.25 \cdot 10^{-4}$.
This is much stronger in this model than the bounds on Higgs compositeness from the Higgs couplings to the SM particles set by the LHC data~\cite{deBlas:2018tjm} ($ \xi \lesssim 0.1 $), and the EW precision measurements~\cite{Cacciapaglia:2020kgq} ($ \xi \lesssim 0.04$). On the contrary, as we expect $m_{DM} \sim f_R$, the values seem compatible with the limit in Eq.~\eqref{DMmass cond1}.
The fine-tuning of one part in $4\cdot 10^{3}$ should not discourage the study of this model, as it represents a huge improvement over the fine-tuning in the SM, which for the Higgs mass amounts to one part in $10^{34}$ against the Planck scale. 
As already discussed, due to the large DM masses, additional tuning is needed in Eq.~\eqref{eq: OmegaDM/OmegaB fundamental} to obtain the correct relic density via the asymmetric production.

Another valid possibility consists in tuning the mass of the DM to be $m_{DM} \ll f_R$. Assuming $\vert X/B\vert = \mathcal{O}(1)$ in Eq.~\eqref{eq: OmegaDM/OmegaB fundamental} (and $T_F \gg m_{DM}$), saturating the relic density would require $m_{DM} \approx 1$~GeV. 
This low mass can be achieved by properly tuning some couplings in the effective potential in Eq.~\eqref{eq:Left}. As the DM pNGB is nearly massless, this limit is technically natural according to 't Hooft naturalness principle~\cite{tHooft:1979rat} as it reveals the restoration of a global symmetry.\\

In summary, we have presented a family of Techni-Composite Higgs models which employ new strong dynamics to produce both a pNGB Higgs and an asymmetric DM candidate. The key novel ingredient is a Left-Right symmetry and the contemporary presence of a Composite Higgs vacuum in the L-sector and a Technicolor vacuum in the R-sector. 
An illustrative example of the allowed parameter space is shown in Fig.~\ref{fig: Model Constraints}. This shows model-independently the main features of Techni-Composite models: a successful ADM relic density can be obtained for masses above $10$~TeV or for masses around the GeV scale. 
This family of models can thereby naturally explain the observed Higgs mass and DM abundance with a minimum of tuning, although some tuning remains necessary to ensure the correct vacuum alignment and DM mass.

\section*{Acknowledgements}
GC acknowledges partial support from the Labex-LIO (Lyon Institute of Origins) under grant ANR-10-LABX-66 (Agence Nationale pour la Recherche), and FRAMA (FR3127, F\'ed\'eration de Recherche ``Andr\'e Marie Amp\`ere''). 
MTF, WCH, and MR acknowledge partial funding from The Council For Independent Research, grant number DFF 6108-00623. The CP3-Origins centre is partially funded by the Danish National Research Foundation, grant number DNRF90. PS acknowledges support by the Deutsche Forschungsgemeinschaft under Germany’s Excellence Strategy - EXC 2121 “Quantum Universe” - 390833306.

\bibliographystyle{JHEP-2-2}

\bibliography{RTC-LCH.bib}

\clearpage
\onecolumngrid

\include{SupplementaryMaterial_included}

\end{document}

%% file: SupplementaryMaterial_included.tex
\renewcommand{\thesubsection}{S\arabic{subsection}}
\setcounter{section}{0}
\renewcommand{\thefigure}{F\arabic{figure}}
\setcounter{figure}{0}
\renewcommand{\thetable}{T\arabic{table}}
\setcounter{table}{0}
\renewcommand{\theequation}{E\arabic{equation}}
\setcounter{equation}{0}

\begin{center}
	\Large 
\textbf{Supplementary material}
\end{center}

\section{Effective Lagrangian and vacuum alignment}
We consider the scenario with $ \mathcal{R}_R$ real and $\mathcal{R}_L$ pseudoreal of $G_{\rm HC}$, which applies to the model studied in the main text. To describe the general vacuum alignment in the effective Lagrangian of this scenario, we identify an $ \SU(2)_{\rm L}\times \UU(1)_{\rm Y'} $ subgroup in $ \SU(4)_{L} $ by the generators \begin{equation}\begin{aligned}
T_{L{\rm L}}^i &= \begin{pmatrix} \sigma_i & 0 \\  0  & 0 \end{pmatrix}, \quad T_{L{\rm Y'}}^3 = \begin{pmatrix} 0 & 0 \\  0  & -\sigma_3^T \end{pmatrix},
\end{aligned} \end{equation} and an $ \SU(2)_{\rm R}\times \UU(1)_{\rm Y'} $ subgroup in $ \SU(4)_{R} $ by the generators\begin{equation}\begin{aligned}
T_{R{\rm R}}^i &= \begin{pmatrix} \sigma_i & 0 \\  0  & 0 \end{pmatrix}, \quad T_{R{\rm Y'}}^3 = \begin{pmatrix} 0 & 0 \\  0  & -\sigma_3^T \end{pmatrix},
\end{aligned} \end{equation} where $ \sigma_i $ are the Pauli matrices with $ i=1,2,3 $. Note that the $ \rm Y' $-charge generator is identified as ${\rm Y'} \equiv T^3_{L{\rm Y'}} + T^3_{R{\rm Y'}}$, while the standard hypercharge is given by ${\rm Y} \equiv T_{L{\rm  Y'}}^3 + T_{R{\rm R}}^3 +T_{R{\rm Y'}}^3$ after the breaking of $\SU(2)_{\rm R}$. Furthermore, $T_{L{\rm Y'}}^3$ and $T_{R{\rm Y'}}^3$ are part of global $\SU(2)$ symmetries that define a custodial symmetry in both cosets \cite{Georgi:1984af}. This is required in the $L$-coset to reproduce the correct $Z$/$W$ mass ratio.

The alignment between the extended EW subgroup $ \SU(2)_{\rm L}\times \SU(2)_{\rm R}\times \UU(1)_{\rm Y'} $ and the stability group $ \SP(4)_{L}\times \SO(4)_{R} $ can then be conveniently parameterized by one misalignment angle, $ \theta_{L} $. To do so, we  identify the vacua that leave the $ \SU(2)_{\rm L}\times \UU(1)_{\rm Y} $ symmetry intact, $ E_{L}^\pm $, and the one breaking $ \SU(2)_{\rm L}\times \UU(1)_{\rm Y} $ to $ \UU(1)_{\rm EM} $, $ E_{L}^B $. In the $R$-coset, there is no vacuum that preserves $ \SU(2)_{\rm R}\times \UU(1)_{\rm Y'}$, hence we define the one breaking the gauge group to $\UU(1)_{\rm Y}$, $ E_{R}^B$. They are given in term of two-index $\SU(4)$ matrices as:
\begin{equation}\begin{aligned}
E_L^\pm &= \begin{pmatrix} i\sigma_2 & 0 \\  0  & \pm i\sigma_2  \end{pmatrix}, \quad E_L^B = \begin{pmatrix} 0 & \mathbb{1}_2 \\ - \mathbb{1}_2  & 0 \end{pmatrix},   \\ E_R^B &= \begin{pmatrix} 0 & \mathbb{1}_2 \\  \mathbb{1}_2  & 0 \end{pmatrix}.
\end{aligned} \end{equation} 
Either choice of $ E_L^\pm $ is equivalent \cite{Galloway:2010bp}, and in this letter we have chosen $ E_L^- $. The true $ \SU(4)_{L} $ vacuum can be written as a linear combination of the above vacua, $ E_{L}(\theta_L) = c_{\theta_{L}}E_{L}^- + s_{\theta_{L}}E_{L}^B $ (a CH vacuum), while the vacuum of the $ \mathbb{S}_R $ sector is $ E_{R}\equiv E_R^B $ (a TC-like vacuum). We use the short-hand notations $ s_x\equiv \sin x $, $ c_x\equiv \cos x $ and $ t_x\equiv \tan x $ throughout.

The Goldstone excitations around the vacuum are then parameterized by \begin{align}
\begin{split}
\Sigma_L(x) &= \exp\left[ 2\sqrt{2}\, i \left( \frac{\Pi_L(x) }{f_L}+\frac{\Pi_\Theta(x)}{f_\Theta}\right)\right] E_L(\theta_L), \\
\Sigma_R(x) &= \exp\left[ 2\sqrt{2}\, i \left( \frac{\Pi_R(x) }{f_R}-\frac{ \Pi_\Theta(x)}{f_\Theta}\right)\right] E_R, 
\end{split}
\end{align} 
where the pion matrices $\Pi_x$ are defined as
\begin{align}
\begin{split}
&\Pi_{L}(x)=\sum^5_{i=1} \Pi_L^i X_L^i  \in G_{L}/H_{L}, \\&
\Pi_{R}(x)=\sum^9_{a=1} \Pi_R^a X_R^a  \in G_{R}/H_{R}, \\&
\Pi_\Theta(x)=\Theta(x)\ \frac{\mathbb{1}_4}{4} \in \UU(1)_\Theta/\varnothing\,.
\end{split}
\end{align} 
The last one encodes the diagonal pNGB state $ \Theta $ associated with the broken global symmetry group $ \UU(1)_\Theta $, acting on the two fermion representations and having no gauge anomaly with the HC group. 
The matrices $ X_L^i $ are the $ \theta_L $-dependent  broken generators of $ \SU(4)_{L} $, while $X_R^a$ are the broken generators of $ \SU(4)_{R} $. While in general the decay constants are different, for simplicity from now on we will assume $ f\equiv f_{L,R}=f_\Theta $.

In the $ \mathbb{S}_L $ sector, we identify the would-be Higgs boson as $ h\equiv \Pi_L^4 \sim c_{\theta_L}(\overline{U}U+\overline{D}D)+s_{\theta_L}{\rm Re}\ U^T\mathcal{C}D $ and the singlet pNGB as $ \eta\equiv  \Pi_L^5 \sim {\rm Im}\ U^T\mathcal{C}D  $, while the remaining three $ \Pi_L^{1,2,3} $ are exact Goldstones eaten by the massive $ W^\pm $ and $ Z $. In the $ \mathbb{S}_R $ sector, we identify the complex isotriplet scalars, \begin{align}
\begin{split}
&\Pi_{CS}^0\equiv \frac{\Pi_R^8+i\Pi_R^9}{\sqrt{2}}\sim C^T \mathcal{C}S, \\ 
&\Pi_{CC}^+\equiv \frac{\Pi_R^4+i\Pi_R^5+\Pi_R^6+i\Pi_R^7}{2}\sim C^T \mathcal{C}C, \\
&\Pi_{SS}^-\equiv \frac{\Pi_R^4+i\Pi_R^5-\Pi_R^6-i\Pi_R^7}{2}\sim S^T \mathcal{C}S, 
\end{split}
\end{align} where $ \Pi_{CS,\overline{CS}}^0 $ is identified as the DM candidate. The remaining three, $ \Pi_R^{1,2,3} $, are exact Goldstones eaten by the massive $ W_{\rm R}^\pm $ and $ Z_{\rm R} $. Note that, following Ref.~\cite{Ryttov:2008xe}, we have used Dirac spinors to indicate the hyperfermions, combining the $\SU(2)_{\rm L,R}$ doublet and singlet Weyl spinors. 

Below the condensation scale $ \Lambda_{\rm HC}\sim 4\pi f$, the effective Lagrangian is given by
\begin{equation}\begin{aligned}
\mathcal{L}_{\rm  eff}&=\mathcal{L}_{\rm kin}-V_{\rm eff}.
\label{eq:total effective Lagrangian}
\end{aligned} \end{equation} 
The kinetic part of the Lagrangian is given by \begin{equation}\begin{aligned}
\label{eq:kinLag}
\mathcal{L}_{\mathrm{kin}}=\frac{f^2}{8}  \text{Tr} [D_{\mu}\Sigma_L^{\dagger}D^{\mu}\Sigma_L]+\frac{f^2}{8}  \text{Tr} [D_{\mu}\Sigma_R^{\dagger}D^{\mu}\Sigma_R].
\end{aligned} \end{equation} with \begin{equation}\begin{aligned} 
&D_\mu\Sigma_{L/R}=\partial_\mu \Sigma_{L/R}-i(G_{L/R,\mu}\Sigma_{L/R}+\Sigma_{L/R}G_{L/R,\mu}^T), \\
& G_{L/R,\mu}=g_{\rm L/R}W_{{\rm L/R},\mu}^i T_{\rm L/R}^i +g_{\rm Y'}B_\mu' T_{L{\rm Y'}/R{\rm Y'}}^3,  \nonumber
\end{aligned} \end{equation} where $ g_{\rm L,R,Y'} $ are the gauge couplings of $ \SU(2)_{\rm L} $, $ \SU(2)_{\rm R} $ and $ \UU(1)_{\rm Y'} $, respectively. Here the hypercharge coupling is given by $ g_{\rm Y}^{-2}=g_{\rm Y'}^{-2}+g_{\rm R}^{-2} $. Henceforth, the minimal value $g_{\rm R}$ can acquire is $g_{\rm R}^{\rm min} = g_{\rm Y}$.

At leading order, each source of symmetry breaking contributes independently to the effective potential in Eq.~(\ref{eq:total effective Lagrangian}): \begin{align} 
V_{\rm eff}\supset V_{\rm gauge}+V_{\rm top}+V_{\rm m}+V_{\rm B}.
\end{align} Here the EW gauge interactions in Eq.~(\ref{eq:kinLag}) yield the gauge loop contributions to the potential $ V_{\rm eff} $: 
\begin{equation}
\begin{aligned}
V_{\text{gauge}}= &-C_L f^4 \left\{ g_{\rm L}^2 \sum_i \text{Tr}\left[ T^i_{L{\rm L}}\Sigma_L (T^i_{L{\rm L}}\Sigma_L)^*\right] + g_{\rm Y'}^2  \text{Tr}\left[ T_{L{\rm Y'}}^3\Sigma_L (T_{L{\rm Y'}}^3\Sigma_L)^*\right] \right\} \\
&+ C_R f^4 \left\{ g_{\rm R}^2 \sum_i \text{Tr}\left[ T^i_{R{\rm R}}\Sigma_R (T^i_{R{\rm R}}\Sigma_R)^*\right]+ g_{\rm Y'}^2 \text{Tr}\left[ T_{R{\rm Y'}}^3\Sigma_L (T_{R{\rm Y'}}^3\Sigma_L)^*\right] \right\},
\end{aligned}\label{Gauge loop contribution (full)}
\end{equation}
where $ C_{L,R} $ encode non-perturbative low energy constants. The top loop contributions arising from the six-fermion operator
\begin{align} \label{eq:topmass supplementary}
\frac{\xi_t}{\Lambda_t^5} (Q_L^T P_L Q_L)(Q_R^T P_R Q_R) q_{{\rm L},3} q_{{\rm R},3}  + {\rm h.c.}
\end{align}	
yield the effective potential contribution:\begin{equation}\begin{aligned}
V_{\rm top}&=-\frac{1}{4}C_t y_t^2 f^4  \Big\{ \left|\Tr[P_L^1\Sigma_L]\right|^2 \left|\Tr[P_R^2\Sigma_R]\right|^2 \\ &+ \Tr[P_L^1\Sigma_L] \Tr[P_R^2\Sigma_R]\Tr[P_L^2\Sigma_L^\dagger] \Tr[P_R^1\Sigma_R^\dagger] \\& +(\text{L}\leftrightarrow \text{R})\Big\},
\end{aligned}
\end{equation} where $ C_t $ is a non-perturbative coefficient for the top loop, $ y_t $ (proportional to $ (\Lambda_{\rm HC}/\Lambda_t)^5 \xi_t  $) is identified by the top Yukawa coupling and the projectors, 
\begin{equation}\begin{aligned}
&(P_{L,R}^1)_{ij}=\frac{1}{2}(\delta_{1i}\delta_{3j}\pm\delta_{3i}\delta_{1j}), \qquad  (P_{L,R}^2)_{ij}=\frac{1}{2}(\delta_{2i}\delta_{4j}\pm\delta_{4i}\delta_{2j}),
\end{aligned}
\end{equation} select the components of $ Q_{L,R}^T Q_{L,R} $ that transform as doublets of $ \SU(2)_{\rm L,R} $. Moreover, the explicit hyper-fermion bilinear mass terms yield potential contributions:
\begin{align}
V_{\rm m} = 2\pi Z_L  \Tr[M_L \Sigma_L^\dagger ]+\rm h.c., \label{Effective mass Lagrangian VL and Dirac masses}
\end{align} where $ M_L = {\rm diag}(m_{1}\epsilon,-m_{2}\epsilon) $ is the mass matrix of the $\mathbb{S}_L$ hyper-fermions. Finally, it is relevant to consider possible four-hyper-fermion operators in the $\mathbb{S}_R$ sector of the form $ Q_R Q_R Q_R Q_R $ 
which yield potential contributions~\cite{Foadi:2007ue}:
\begin{align}
V_{\rm B}&=C_B g_B^2 f^4  \Tr[B\Sigma_{R}^{\dagger}B\Sigma_{R}].
\label{Effective mass Lagrangian}
\end{align} The coefficients $ C_{L,R} $, $ C_{t} $, $ Z_L $ and $ C_B $ in Eq.~(\ref{Gauge loop contribution (full)})-(\ref{Effective mass Lagrangian}) are $ \mathcal{O}(1) $ form factors that can be computed on the lattice~\cite{Arthur:2016ozw}.
At leading order, the effective potential for the misalignment angle is given by
\begin{equation}
\begin{aligned} 
V_{\rm eff}^0=&-8\pi f^3 Z_L m_{UD} c_{\theta_L}- f^4 C_L  \widetilde{g}_L^2c_{\theta_L}^2 -f^4 C_t y_t^2 s_{\theta_L}^2  ,
\end{aligned} 
\end{equation} where $ m_{UD}\equiv m_1+m_2 $ and $  \widetilde{g}_L^2\equiv (3 g_{\rm L}^2 +g_{\rm Y'}^2)/2 $. By minimizing the above potential, $ \partial_{\theta_L} V_{\rm eff}^0=0 $, we obtain \begin{equation}\begin{aligned}
&c_{\theta_L} =\frac{4\pi Z_L m_{UD}}{f(C_t y_t^2- C_L\widetilde{g}_L^2)},  
\end{aligned}\end{equation} where the $ G_L/H_L $ part of the vacuum is aligned in a composite Higgs direction ($ 0<c_{\theta_L}<1 $), while the $ G_R/H_R $ part of the vacuum remains in a TC direction. The conditions for this vacuum alignment to be a stable minimum in the presence of these terms are $ \partial_{\theta_L\theta_L}V_{\rm eff}^0>0 $ and that all the squared masses of the pNGBs are positive. However, the condition $ \partial_{\theta_L\theta_L}V_{\rm eff}^0>0 $ is fulfilled by requiring $ m_h^2>0 $ due to the fact that $ \partial_{\theta_L\theta_L}V_{\rm eff}^0 = f^2 m_h^2  $. Therefore, we need that all the pNGBs have positive squared masses. The Higgs and $ \eta $ masses are \begin{equation}\begin{aligned}
&m_h^2 =2(C_t y_t^2 - C_L \widetilde{g}_L^2 )v_{\rm EW}^2,\quad \quad m_\eta^2 = m_h^2 /s_{\theta_L}^2,
\end{aligned}\end{equation} while the masses of the complex isotriplet composite scalars, $ \Pi_{CS,\overline{CS}}^0,\Pi^\pm_{CC},\Pi_{SS}^\pm $, in the $ \mathbb{S}_R $ sector are  \begin{equation}\label{eq:massesDM}\begin{aligned}
&m_{\Pi_{CS}^0}^2 = 2( C_Bg_B^2  - C_R \widetilde{g}_R^2 + C_t y_t^2 s_{\theta_L}^2)f^2, \\ 
&m_{\Pi_{CC}^\pm}^2=m_{\Pi_{SS}^\pm}^2 = m_{\Pi_{CS}^0}^2+2C_R g_{Y'}^2 f^2,
\end{aligned}\end{equation} where $  \widetilde{g}_R^2\equiv (-g_{\rm R}^2+g_{\rm Y'}^2)/2 $. To obtain a stable vacuum, we therefore need $ m_h^2 >0 $ and $ m_{\Pi_{CS}^0}^2>0 $. 
Following the common lore that the top loops tend to break the electroweak symmetry while gauge loops preserve it, we assume that the form factors $C_{L,R,t} > 0$. Hence, the masses squared are always positive, as long as the top loops dominate, as expected as the Yukawa coupling is larger than the gauge ones. Furthermore, in the $R$-coset, we find that $\widetilde{g}_R^2 \geq 0$ for $g_{\rm R} \geq \sqrt{2} g_{\rm Y}$, while it becomes positive for $g_{\rm Y} < g_{\rm R} < \sqrt{2} g_{\rm Y}$.   In the latter range, it tends to cancel the contribution of the top. As $m_{\Pi_{CS}}^0 = m_{DM}$, the magnitude of the Dark Matter candidate mass crucially depends on the value of these coefficients. In general, the first Eq.~\eqref{eq:massesDM} implies that the DM mass is of the order of the pNGB decays constants, i.e. $m_{DM} \sim f$. Small masses, of the order of GeV, could be obtained if a tuned cancellation is enacted. This could happen if the gauge contribution is large and positive, for $g_{\rm R} \gtrsim g_{\rm Y}$, or for $C_B < 0$.

As discussed in the main text, for light DM mass the annihilation is dominantly into the light pNGB associated to the $\UU(1)_\Theta$ symmetry. If either of the vector-like masses $ m_1 $ or $m_2$ are vanishing, the pNGB $ \Theta $ state mixes with the $ \mathbb{S}_L $ pNGB $ \eta $ state, resulting in that the mass eigenstate $ \widetilde{\Theta} $, consisting mostly of $ \Theta $, is massless while $ \widetilde{\eta} $ has a mass of order $ f $:\begin{equation}\begin{aligned}
&m_{\widetilde{\Theta}}^2=0,\quad \quad m_{\widetilde{\eta}}^2=\frac{1}{4} \frac{m_h^2}{s_{\theta_L}^2} (5+c_{2\theta_L})\approx  \frac{3m_h^2}{2s_{\theta_L}^2}.
\end{aligned}\end{equation} However, the $ \widetilde{\Theta} $ state can achieve a small mass from its mixing with a $ \Theta' $ state corresponding to the $ \UU(1)$ symmetry which is quantum anomalous. In addition, the mass of $ \Theta' $ is generated by instanton effects related to the $ \UU(1) $ anomaly~\cite{Belyaev:2016ftv}. 

Thus, for DM masses below the $W_{\rm R}$ mass, the dominant annihilation channel is $ \Pi_{X} \overline{\Pi}_{X} \to \widetilde{\Theta}\widetilde{\Theta}$ with the coupling $ \lambda_{XX \widetilde{\Theta} \widetilde{\Theta}} $ given by\begin{equation}\begin{aligned}
&\mathcal{L}_{\rm eff}\supset \lambda_{XX\widetilde{\Theta}\widetilde{\Theta}}\Pi_{CS}^0\Pi_{\overline{CS}}^0\widetilde{\Theta}\widetilde{\Theta}
\end{aligned}\end{equation} with \begin{equation}\begin{aligned}
\lambda_{XX\widetilde{\Theta}\widetilde{\Theta}}=&4C_t y_t^2 s_{\theta_L}^2 \frac{c_{\theta_L}}{\sqrt{2+c_{\theta_L}^2}}\approx \frac{4}{\sqrt{3}}C_t y_t^2 \left(\frac{v_{\rm EW}}{f}\right)^2 \equiv \lambda_0  \left(\frac{v_{\rm EW}}{f}\right)^2.  \nonumber
\end{aligned}\end{equation} 
This implies that $\lambda_0 \sim \frac{4}{\sqrt{3}}C_t y_t^2$, which could be an order 1 number.\\

In the scenario with $ \mathcal{R}_R = F $ (fundamental, pseudo-real) of $ G_{\rm HC}=\SP(2N_{\rm HC}) $, the DM candidate is identified by a complex isosinglet, $ \Pi_{CS,\overline{CS}} $, where its mass is given by \begin{equation}\begin{aligned}
&m_{\Pi_{CS}}^2 = 2( C_Bg_B^2  - C_R \widetilde{g}_R^2 + C_t y_t^2 s_{\theta_L}^2)f^2,
\end{aligned}\end{equation} where $ \widetilde{g}_R^2=(3g_{\rm R}^2+g_{\rm Y'}^2)/2 $ which is always positive. Due to the fact that $ s_{\theta_L}\ll 1 $, $ m_{\Pi_{CS}}^2 $ is negative when $ C_B g_B^2=0 $ and therefore a small DM mass, $ m_{\Pi_{CS}}\ll f $, can be achieved by tuning $ C_B g_B^2 $ to a certain value of order unity. Furthermore, in the scenarios with the top mass arising from PC operators, the DM mass can also be tuned by the term $ C_B g_B^2 $ to small values in both scenarios.  These models are inspired by the work in Ref.~\cite{Ferretti:2013kya}.


\section{Examples of theories featuring the Techni-Composite Higgs mechanism}

In the main body of the article, we studied in detail one specific model based on a gauge symmetry $\SP(2N)_{\rm HC}$ and with fermions in two different representations. However, the same mechanism can be found in many other models, with different possibilities for the $ L $ and $ R $-cosets, as listed in Tables~\ref{table:Lcosets} and~\ref{table:Rcosets}, respectively.  Here we list the quantum numbers of the HC fermions needed to obtain the minimal cosets, for the three classes of HC representations: pseudo-real, real and complex.

\begin{table}[htb]
	\centering
	\begin{tabular}{cccccc}
		\hline
		&$G_{\rm HC}$		&$\SU(2)_{\rm L}$	&$\SU(2)_{\rm R}$	&$ \UU(1)_{\rm Y'} $	&$\UU(1)_{TB,L}$\\
		\hline
		\multicolumn{6}{c}{pseudo-real} \\ \hline
		$ (U,D) $	&$\mathcal{R}_{L}$				&$ \Box $		&1			&0				& 0\\
		$ \widetilde{U} $	&$\mathcal{R}_{L}$					&1			&1			&$ -1/2	 $		& 0	\\
		$ \widetilde{D} $	&$\mathcal{R}_{L}$				&1			&1			&$ +1/2 $			& 0	\\
		\hline
		\multicolumn{6}{c}{complex} \\ \hline
		$ (U,D) $	&$\mathcal{R}_{L}$				&$ \Box $		&1			&0				& 1\\
		$ \widetilde{U} $	&$\mathcal{R}_{L}$					&1			&1			&$ -1/2	 $		& 1	\\
		$ \widetilde{D} $	&$\mathcal{R}_{L}$				&1			&1			&$ +1/2 $			& 1	\\
		$ (U^c,D^c) $	&$\mathcal{R}_{L}^\ast$				&$ \Box $		&1			&0				& -1\\
		$ \widetilde{U}^c $	&$\mathcal{R}_{L}^\ast$					&1			&1			&$ +1/2	 $		& -1	\\
		$ \widetilde{D}^c $	&$\mathcal{R}_{L}^\ast$				&1			&1			&$ -1/2 $			& -1	\\
		
		\hline
		\multicolumn{6}{c}{real} \\ \hline
		$ (U,D) $	&$\mathcal{R}_{L}$				&$ \Box $		&1			&1/2				& 0\\
		$ (\widetilde{U},\widetilde{D}) $	&$\mathcal{R}_{L}$		&$ \Box $		&1			&-1/2				& 0\\
		$ X $	&$\mathcal{R}_{L}$					&1			&1			& 0			& 0	\\ \hline
	\end{tabular}
	\caption{Fermion field content and their charges for the minimal L-cosets.}
	\label{table:Lcosets}
\end{table}

\begin{table}[htb]
	\centering
	\begin{tabular}{ccccccc}
		\hline
		&$G_{\rm HC}$		&$\SU(2)_{\rm L}$	&$\SU(2)_{\rm R}$	&$ \UU(1)_{\rm Y'} $	&$\UU(1)_X$ & $\UU(1)_{TB,R}$\\
		\hline
		\multicolumn{7}{c}{pseudo-real} \\ \hline
		$ (C,S) $	&$\mathcal{R}_{R}$						&1			&$ \Box $		&0				&$ +1$	& 0\\
		$ \widetilde{C} $	&$\mathcal{R}_{R}$					&1			&1			&$ -1/2 $			&$ -1$ & 0\\
		$ \widetilde{S} $	&$\mathcal{R}_{R}$					&1			&1			&$ +1/2 $			&$ -1$& 0\\		\hline
		\multicolumn{7}{c}{complex} \\ \hline
		$ (C,S) $	&$\mathcal{R}_{R}$						&1			&$ \Box $		&0				&$ +1$	& 1\\
		$ \widetilde{C} $	&$\mathcal{R}_{R}^\ast$					&1			&1			&$ -1/2 $			&$ -1$ & $-1$\\
		$ \widetilde{S} $	&$\mathcal{R}_{R}^\ast$					&1			&1			&$ +1/2 $			&$ -1$ & $-1$\\		\hline
		\multicolumn{7}{c}{real} \\ \hline
		$ (C,S) $	&$\mathcal{R}_{R}$						&1			&$ \Box $		&0				&$ +1$ & 0	\\
		$ \widetilde{C} $	&$\mathcal{R}_{R}$					&1			&1			&$ -1/2 $			&$ -1$ & 0\\
		$ \widetilde{S} $	&$\mathcal{R}_{R}$					&1			&1			&$ +1/2 $			&$ -1$ & 0\\ \hline
	\end{tabular}
	\caption{Fermion field content and their charges for the minimal R-cosets.}
	\label{table:Rcosets}
\end{table}

Finally, in Table~\ref{tab:examples} we provided some examples of gauge-fermion theories generating various combinations for the $L$ and $R$-cosets. In the cases with $\mathcal{R}_L = \mathcal{R}_R$, the global symmetry is extended to a single simple group that contains the $L$ and $R$ sub-cosets

\begin{table}[htb]
	\centering
	\begin{tabular}{|l|c|c|c|c|l|}
		\hline
		$G_{\rm HC}$ & $\mathcal{R}_L$ & dim($\mathcal{R}_L$) & $\mathcal{R}_R$ & dim($\mathcal{R}_R$) & Annotations\\ \hline 
		\multicolumn{6}{c}{\phantom{\Big(} $\SU(4)/ \SP(4)_L \otimes \SU(4)/\SO(4)_R \otimes \UU(1)_\Theta/\varnothing$} \\ \hline
		$\SP(2N)_{\rm HC}$ & $F$ & $2N$ & $A$ & $N (2N-1)-1$ & $N$ odd to avoid Witten anomalies \\
		$\SP(2N)_{\rm HC}$ & $F$ & $2N$ & $Adj$ & $N (2N+1)$ & $N$ even to avoid Witten anomalies \\
		\hline
		\multicolumn{6}{c}{\phantom{\Big(} $\SU(4)^2/\SU(4)_L \otimes \SU(4)/\SO(4)_R \otimes \UU(1)_\Theta/\varnothing \otimes \UU(1)_{TB, L}$} \\ \hline
		$\SU(4)_{\rm HC}$ & $F$ & $4$ & $A_2$ & $6$ & - \\
		$\SO(10)_{\rm HC}$ & $Spin$ & $16$ & $F$ & $10$ & - \\
		\hline
		\multicolumn{6}{c}{\phantom{\Big(} $\SU(4)^2/\SU(4)_L \otimes \SU(2)^2/\SU(2)_R \otimes \UU(1)_\Theta/\varnothing \otimes \UU(1)_{TB,L} \otimes \UU(1)_{TB,R}$} \\ \hline
		$\SU(5)_{\rm HC}$ & $F$ & $5$ & $A_2$ & $10$ & - \\
		\hline
		\multicolumn{6}{c}{\phantom{\Big(} $\SU(5)/\SO(5)_L \otimes \SU(4)/\SO(4)_R \otimes \UU(1)_\Theta/\varnothing$} \\ \hline
		$\SO(7)_{\rm HC}$ & $F$ & $7$ & $Spin$ & $8$ & - \\
		\hline
		\multicolumn{6}{c}{\phantom{\Big(} $\SU(5)/\SO(5)_L \otimes \SU(4)/\SP(4)_R \otimes \UU(1)_\Theta/\varnothing$} \\ \hline
		$\SP(2N)_{\rm HC}$ & $A$ & $N (2N-1)-1$ & $F$ & $2N$ & $N \geq 2$ \\
		\hline
		\multicolumn{6}{c}{\phantom{\Big(} $\SU(5)/\SO(5)_L \otimes \SU(2)^2/\SU(2)_R \otimes \UU(1)_\Theta/\varnothing \otimes \UU(1)_{TB,R}$} \\ \hline
		$\SU(4)_{\rm HC}$ & $A$ & $6$ & $F$ & $4$ & - \\
		$\SO(10)_{\rm HC}$ & $F$ & $10$ & $Spin$ & $16$ & - \\
		\hline
		\multicolumn{6}{c}{\phantom{\Big(} $\SU(8)/\SP(8)$} \\ \hline
		$\SP(2N)_{\rm HC}$ & $F$ & $2N$ & $F$ & $2N$ & - \\
		\hline
		\multicolumn{6}{c}{\phantom{\Big(} $\SU(9)/\SO(9)$} \\ \hline
		$\SU(4)_{\rm HC}$ & $A$ & $6$ & $A$ & $6$ & - \\
		$\SO(10)_{\rm HC}$ & $F$ & $10$ & $F$ & $10$ & - \\
		\hline
		\multicolumn{6}{c}{\phantom{\Big(} $\SU(6)^2/\SU(6) \otimes \UU(1)_{TB}$} \\ \hline
		$\SU(N)_{\rm HC}$ & $F$ & $N$ & $F$ & $N$ & - \\
		$\SU(5)_{\rm HC}$ & $A$ & $10$ & $A$ & $10$ & - \\
		\hline
	\end{tabular}
	\caption{\label{tab:examples} Examples of gauge fermion theories leading to the Techni-Composite Higgs models.}
\end{table}

\clearpage
\section{Asymmetry sharing}
We here in detail show  how the asymmetric dark matter relic can be calculated \cite{Harvey:1990qw,Ryttov:2008xe}.
First, note that at high temperatures the particle number density $ n_+ $ and the anti-particle number density $ n_- $ of a given species are given by
\begin{align}
n_\pm =  g \int \frac{d^3k}{(2\pi)^3}\frac{g}{e^{(E\mp\mu)\beta}-\eta} \qq{with} \begin{aligned}
\eta &= +1 \text{ for bosons} \\[-1ex]
\eta &= -1 \text{ for fermions}
\end{aligned}
\end{align}
where $ g $ is the number of internal degrees of freedom, $ \mu $ is the chemical potential of the particle species and $ \beta = 1/T$ (with $ k_B = 1 $). 
At the freeze-out temperature of sphalerons, $ T_F $, we have $ \mu/T_F\ll 1$ such that the difference in particle numbers of a given species is given by
\begin{gather}
n = n_+ - n_- = gT_F^3 \frac{\mu}{T_F}\frac{\sigma(m/T_F)}{6}, \label{eq:assymetric number density}
\end{gather}
which reveals that the chemical potentials are the relevant quantities.
Here the statistical suppression factor $ \sigma $ is
\begin{gather}
\sigma(m/T_F)=\begin{cases}
\frac{6}{4\pi^2}\int_{0}^{\infty} dx x^2 \cosh[-2](\frac{1}{2}\sqrt{x^2+(m/T_F)^2})	&\text{for fermions}\\
\frac{6}{4\pi^2}\int_{0}^{\infty} dx x^2 \sinh[-2](\frac{1}{2}\sqrt{x^2+(m/T_F)^2})	&\text{for bosons.}
\end{cases} 
\label{Eq:statfunc}
\end{gather}
The statistical factor $ \sigma(m/T_F) $ is normalized such that
\begin{gather}
\lim_{m/T_F\to 0} \sigma(m/T_F) = \begin{cases}
1 &\text{for fermions,} \\
2 &\text{for bosons,}
\end{cases} 
\end{gather}
while $\sigma(m/T_F)\simeq2 (m/2\pi T_F)^{3/2} e^{-m/T_F}$ for large $m/T_F\gg 1$. 
In therms of chemical potentials, the ratio of DM and baryon energy densities can be expressed as
\begin{gather}
\frac{\Omega_{DM}}{\Omega_B} = \frac{\frac{1}{2}m_{DM} }{m_P}\left|\frac{\mu_{X}}{\mu_B}\right|\sigma\left(\frac{m_{DM}}{2 T_F}\right),
\end{gather}
where we neglected the $ \sigma(m / 2 T_F) $ for the SM fermions. The task is then to calculate the total chemical potential $ \mu_{X} $ of all fermions charged under $\UU(1)_X$  and the total chemical potential of all baryons $ \mu_B $


To study the chemical potentials we adopt the we adopt the notation of Ryttov-Sannino \cite{Ryttov:2008xe} and Harvey-Turner \cite{Harvey:1990qw}. The equilibrium condition from sphalerons, as written in this convention, was already given in the main text. Additionally, the 6-fermion operators, given by Eq.~\eqref{eq:topmass supplementary}, lead to equilibrium conditions of the form
\begin{align}
\begin{aligned}
\mu_{UL}-\mu_{UR}+\mu_{SR}-\mu_{SL}		-\mu_{u{\rm L},i}+\mu_{u{\rm R},i} 		&=0,\\
\mu_{UL}-\mu_{UR}+\mu_{SR}-\mu_{SL}		+\mu_{d{\rm L},i}-\mu_{d{\rm R},i} 		&=0,\\
\mu_{UL}-\mu_{UR}+\mu_{SR}-\mu_{SL}		-\mu_{\nu {\rm L},i}+\mu_{\nu{\rm R},i}&=0,\\
\mu_{UL}-\mu_{UR}+\mu_{SR}-\mu_{SL}		+\mu_{e{\rm L},i}-\mu_{e{\rm R},i} 		&=0,
\end{aligned}\label{eq:6-fermion block 1}
\end{align}
and
\begin{align}
\begin{aligned}
\mu_{DL}-\mu_{DR}+\mu_{CR}-\mu_{CL}		-\mu_{u{\rm R},i}+\mu_{u{\rm L},i}		&=0,\\
\mu_{DL}-\mu_{DR}+\mu_{CR}-\mu_{CL}		+\mu_{d{\rm R},i}-\mu_{d{\rm L},i}		&=0,\\
\mu_{DL}-\mu_{DR}+\mu_{CR}-\mu_{CL}		-\mu_{\nu{\rm R},i}+\mu_{\nu{\rm L},i}&=0,\\
\mu_{DL}-\mu_{DR}+\mu_{CR}-\mu_{CL}		+\mu_{e{\rm R},i}-\mu_{e{\rm L},i}		&=0,\\
\end{aligned}\label{eq:6-fermion block 2}
\end{align}
where the index $ _i $ refers to the three generations of SM fermions, each of which has equilibrium conditions of this form. However, depending on the temperature, only some generations of quarks and leptons have efficient 6-fermion operators. As was pointed out in the letter, a fermion $\psi$, receiving its mass $m_\psi$ from such an interaction is in equilibrium at a temperature $ T $ if~\cite{Davidson:2008bu, Deppisch:2017ecm}
\begin{gather}
2.3\times 10^4\times \sqrt{\frac{10\text{ TeV}}{f}}\frac{m_\psi}{v_{\rm EW}} \gtrsim \left(\frac{f}{T}\right)^{9/2}. \label{eq:in equilibrium condition}
\end{gather}
We assume that the conditions corresponding to \eqref{eq:6-fermion block 1} and \eqref{eq:6-fermion block 2} apply for any fermions for which the condition \eqref{eq:in equilibrium condition} is satisfied for all participating species at the condensation temperature $ T=f $.

Furthermore, thermal equilibrium in the electroweak interactions imply the following conditions:
\begin{align}
\begin{aligned}\label{EW interactions equilibrium equations}
\mu_{u{\rm L},i}&=\mu_{d{\rm L},i}+\mu_{W\rm L}, & \mu_{u{\rm R},i}&=\mu_{d{\rm R},i}+\mu_{W\rm R}, \\
\mu_{e{\rm L},i}&=\mu_{\nu{\rm L}, i}+\mu_{W\rm L}, & \mu_{e{\rm R},i}&=\mu_{\nu{\rm R}, i}+\mu_{W\rm R}, \\ 
\mu_{DL}&=\mu_{UL}+\mu_{W\rm L}, & \mu_{SL}&=\mu_{CL}+\mu_{W\rm R}. \\ 
\end{aligned}
\end{align}
The equilibrium is constrained by conditions on the net charge of plasma. To quantify this, consider the eigenvalues of third the components of the isospins of $ \SU(2)_{\rm L,R} $, which are
\begin{align}
T_{L\rm L}^3 = &3\times \sum_i\left( \frac{1}{2}\mu_{u{\rm L},i}-\frac{1}{2}\mu_{d{\rm L},i}\right) + \sum_i\left( \frac{1}{2}\mu_{\nu{\rm L}, i}-\frac{1}{2}\mu_{e{\rm L}, i} \right)-4\mu_{W\rm L}+2\left( \frac{1}{2}\mu_{UL}-\frac{1}{2}\mu_{DL} \right),\\
T_{R\rm R}^3 = &3\times \sum_i\left( \frac{1}{2}\mu_{u{\rm R},i}-\frac{1}{2}\mu_{d{\rm R},i}\right) + \sum_i\left( \frac{1}{2}\mu_{\nu{\rm R}, i}-\frac{1}{2}\mu_{e{\rm R}, i} \right)-4\mu_{W\rm R}+2\left( \frac{1}{2}\mu_{CR}-\frac{1}{2}\mu_{SR} \right).
\end{align}
The overall electric charge, $ Q = T_{L\rm L}^3+T_{R\rm R}^3+Y' $, is then
\begin{align}
Q = &3\times \sum_i\left( \frac{2}{3}\mu_{u{\rm L},i}+\frac{2}{3}\mu_{u{\rm R},i}-\frac{1}{3}\mu_{d{\rm L},i}-\frac{1}{3}\mu_{d{\rm R},i} \right)+\sum_i\left( -1 \mu_{e{\rm L},i}-1 \mu_{e{\rm R},i}+0\mu_{\nu{\rm  L}, i}+0\mu_{\nu{\rm  R}, i} \right)\notag\\
&+ d(\mathcal{R}_L)\left( \frac{1}{2}\mu_{UL}+\frac{1}{2}\mu_{UR}-\frac{1}{2}\mu_{DL}-\frac{1}{2}\mu_{DR} \right) + d(\mathcal{R}_R)\left( \frac{1}{2}\mu_{CL}+\frac{1}{2}\mu_{CR}-\frac{1}{2}\mu_{SL}-\frac{1}{2}\mu_{SR} \right)\\ 
&-4\mu_{W\rm L}-4\mu_{W\rm R}. \notag
\end{align} 
The conditions imposed depend on the type of phase transition. Above the transition, where $ \SU(2)_{\rm L,R} $ are good symmetries, $ T_{ {L\rm L},{R\rm R}}^3 $ must vanish. If the phase transition is sudden, i.e. first order, then this property is assumed to be inherited by the relic such that relevant condition is
\begin{gather}
Q = 0, \quad T_{L \rm L}^3 = 0, \quad T_{R \rm R}^3 = 0.\label{First-order condition,app}
\end{gather}
Below the transition, $ \SU(2)_{\rm L,R} $ are broken, such that $ T_{{L\rm L},{R\rm R}}^3 $ need no longer vanish. If the phase transition is gradual, i.e. second order, then this leads to violation of $ T_{{L\rm L},{R\rm R}}^3 $ conditions. Instead, the VEV drives the chemical potentials of neutral condensates to zero, such that the relevant condition is 
\begin{gather}
\begin{gathered}
Q = 0, \quad \mu_{UL}-\mu_{UR}= 0, \quad \mu_{CR}-\mu_{CL}= 0. \label{Second-order condition,app}
\end{gathered}
\end{gather}
We are ultimately interested in comparing the total baryon and DM densities, which depend on the total chemical potentials of the baryons and technibaryons (charged under $\UU(1)_X$), which are
\begin{align} \label{Baryon,lepton, TB number}
\mu_B &= \sum_i \left(\mu_{u{\rm L},i}+\mu_{u{\rm R},i} +\mu_{d{\rm L},i}, + \mu_{d{\rm R},i} \right), \\
\mu_{X} &= d(\mathcal{R}_R) \left( \mu_{CR}+\mu_{CL}+\mu_{SL}+\mu_{SR} \right).
\end{align}
Solving all the above conditions for $ \mu_{X} $ and $ \mu_{B} $ we find
\begin{gather}
\left|\frac{\mu_{X}}{\mu_B}\right| = 2\left(3+\frac{L}{B}\right), \label{eq:mu ratio}
\end{gather}
if 6-fermion operations are inefficient for at least one generation but efficient for at least one other generation. Since the occupation numbers are proportional to the chemical potentials, we can identify $ \left|\mu_{X}/\mu_B\right| = \left|X/B\right| $. If all generations are in equilibrium, then the lepton number is constrained to $ L=-3B $ so that $ \mu_{X}=0 $. Furthermore, if 6-fermion operators are inefficient for all generations, then $ \mu_{X} $ is unconstrained in the case of first-order transitions while Eq. \eqref{eq:mu ratio} applies for the case of second-order transitions.